\documentclass[epj]{svjour}
\usepackage{graphicx}
\usepackage{amsmath,amssymb}
\usepackage{times}
\usepackage[latin1]{inputenc}

\def\Ham{\mathcal{H}}
\def\ladder{\text{\textsf{L}}}

\begin{document}

\title{Spin 3/2 fermions with attractive interactions in a
  one-dimensional optical lattice: phase diagrams, entanglement
  entropy, and the effect of the trap}

\author{
G. Roux\inst{1} \and 
S. Capponi\inst{2} \and 
P. Lecheminant\inst{3} \and
P. Azaria\inst{4}
}

\institute{
Institute for Theoretical Physics C, RWTH Aachen University, D-52056 Aachen, Germany. \and
Laboratoire de Physique Th\'eorique - IRSAMC, UPS and CNRS, Universit\'e de Toulouse F-31062 Toulouse, France \and
Laboratoire de Physique Th\'eorique et Mod\'elisation, Universit\'e de Cergy-Pontoise, CNRS,  95302 Cergy-Pontoise, France. \and
Laboratoire de Physique Th\'eorique de la Mati\`ere Condens\'ee, Universit\'e Pierre et Marie Curie, CNRS, 75005 Paris, France.
}

\date{\today}

\abstract{We study spin 3/2 fermionic cold atoms with attractive
  interactions confined in a one-dimensional optical lattice.  Using
  numerical techniques, we determine the phase diagram for a generic
  density. For the chosen parameters, one-particle excitations are
  gapped and the phase diagram is separated into two regions: one
  where the two-particle excitation gap is zero, and one where it is
  finite. In the first region, the two-body pairing fluctuations (BCS)
  compete with the density ones. In the other one, a molecular
  superfluid (MS) phase, in which bound-states of four particles form,
  competes with the density fluctuations. The properties of the
  transition line between these two regions is studied through the
  behavior of the entanglement entropy.  The physical features of the
  various phases, comprising leading correlations, Friedel
  oscillations, and excitation spectra, are presented. To make the
  connection with experiments, the effect of a harmonic trap
  is taken into account. In particular, we emphasize the
  conditions under which the appealing MS phase can be realized, and
  how the phases could be probed by using the density profiles and the
  associated structure factor. Lastly, the consequences on the flux
  quantization of the different nature of the pairing in the BCS and
  MS phases are studied in a situation where the condensate is in a
  ring geometry.}

\authorrunning{G. Roux {\it et al.}}
\titlerunning{Spin 3/2 fermions with attractive interactions in one dimension}
\maketitle

\section{Introduction}

Recent experimental progress achieved in trapped ultracold
atomic gases provides a great opportunity for exploring the physics of
strong correlations in clean systems, thanks to the tunability of
interactions using optical lattices and Feshbach resonance. A large
number of interesting phenomena of condensed matter physics and
nuclear physics is then expected to be accessible in the context of
ultracold atomic gases~\cite{Lewenstein2006}. A prominent example is
the observation of the Mott insulator-superfluid quantum phase
transition with cold bosonic atoms in an optical lattice
\cite{Greiner2002}, and its possible fermionic analogue, the Mott
insulator-metallic phase transition, recently investigated in a
two-component Fermi gas~\cite{Jordens2008}. A second breakthrough is
the trapping of a two-component Fermi gas and the study of the
crossover from fermionic superfluidity of Cooper (BCS) pairs to
Bose-Einstein condensation of tightly bounded molecules
\cite{Bloch2007,Giorgini2007,Ketterle2008}.

The superfluid behavior of multicomponent Fermi gases with more than
two hyperfine states might also lead to interesting properties that
have been explored recently
\cite{Koltun1987,Schlottmann1994,Modawi1997,Ho1999,Stepanenko1999,Honerkamp2004,Honerkamp2004a,Wu2005,Lecheminant2005,Kamei2005,Paananen2006,He2006,Bedaque2006,Sedrakian2006,Lee2007,Rapp2007,Cherng2007,Zhai2007,Fukuhara2007,Capponi2008,Rapp2008,Blume2008,Liu2008,Guan2008,Lecheminant2008}.
In particular, the interplay between superfluidity and magnetism,
which stems from the presence of the different internal states, can be
investigated. Experimentally, three component Fermi gases can be
created by trapping the three lowest hyperfine states of $^6$Li atoms
in a magnetic field, or by considering $^{40}$K atoms. In addition,
the magnetic field dependence of the three scattering lengths of
$^6$Li is known experimentally and can be tuned via Feshbach
resonance~\cite{Bartenstein2005} which opens for the experimental
realization of a three-component fermionic lattice model. In fact,
such a degenerate Fermi gas has been realized experimentally very
recently~\cite{Ottenstein2008}, and a four-component Fermi gas could
also be achieved using $^{40}$K atoms~\cite{DeMarco2001}.

The existence of these internal degrees of freedom is expected to give
rise to some exotic superfluid phases. In this respect, a molecular
superfluid (MS) phase might be stabilized where more than two fermions
form a bound state. Such a state might be relevant to several topics
in physics. For instance, the quark model of nuclear matter at low
density describes nucleons as three-fermion bound states. Such a
trionic phase has been found in one-dimensional integrable fermionic
model with three colors \cite{Koltun1987} and its emergence in the
context of three-component ultracold fermions has been discussed
recently \cite{Rapp2007,Capponi2008,Rapp2008,Liu2008,Guan2008}. The
possibility that superfluidity is sustained by a condensate based on
four-fermion bound states (quartet) might be also explored in cold
atomic physics
\cite{Stepanenko1999,Wu2005,Lecheminant2005,Kamei2005,Capponi2008,Lecheminant2008}.
Such a superfluid behavior has already been found in very different
contexts such as nuclear physics for instance, where a four-particle
condensate--the $\alpha$ particle--is known to be favored over
deuteron condensation at low densities \cite{Ropke1998a,Ropke2006}.
Such a quartet condensation can also occur in semiconductors with the
formation of biexcitons~\cite{Nozieres1982}. A quartetting phase,
which stems from the pairing of Cooper pairs, has also been found in a
model of one-dimensional (1D) Josephson junctions~\cite{Doucot2002}
and in four-leg Hubbard ladders~\cite{Chang2007}.

In this paper, we will investigate the low-energy properties of
(hyperfine) spin-3/2 (i.e. four-component) fermionic cold atoms confined in
a one-dimensional optical lattice in light of the possible formation
of a quartetting phase. Due to Pauli's principle, low-energy $s$-wave
scattering processes of spin 3/2 fermionic atoms are allowed in the
singlet and quintet channels, so that the effective Hamiltonian with
contact interactions reads~\cite{Ho1999,Ho1998}
\begin{equation}
\begin{split}
\Ham &= -t \sum_{i,\alpha} [c^{\dagger}_{\alpha,i} c_{\alpha, i+1} + \text{h.c.} ] - \sum_i \mu_i\, n_i\\
& \quad +U_0 \sum_{i} P_{00,i}^{\dagger} P_{00,i} + U_2 \sum_{i,m} P_{2m,i}^{\dagger} P_{2m,i} \;,
\label{eq:sun-hubbardSgen}
\end{split}
\end{equation}
where $c^{\dagger}_{\alpha,i}$ is the fermionic creation operator at
site $i$, in one of the $\alpha = \pm 1/2, \pm 3/2$ hyperfine states.
The on-site density operator is denoted by $n_i = \sum_{\alpha}
c^{\dagger}_{\alpha,i} c_{\alpha,i}$. The chemical potential $\mu_i$
can be uniform (called $\mu$ for grand-canonical Quantum Monte-Carlo
calculations), inhomogeneous in presence of the trap, zero for
DMRG calculation (canonical ensemble). For convenience, the lattice
spacing is set to unity. Singlet and quintet operators in Eq.
(\ref{eq:sun-hubbardSgen}) are defined using Clebsch-Gordan
coefficients
\begin{equation*}
P^{\dagger}_{Jm,i} = \sum_{\alpha \beta}
\langle{Jm}|{\alpha\beta}\rangle c^{\dagger}_{\alpha,i}
c^{\dagger}_{\beta,i}\;.
\end{equation*}
For instance, the spin 3/2 on-site singlet operator reads
$P^{\dagger}_{00,i} = P^{\dagger}_{i} = c^{\dagger}_{3/2,i}
c^{\dagger}_{-3/2,i} - c^{\dagger}_{1/2,i} c^{\dagger}_{-1/2,i} $. A
convenient way to rewrite the Hamiltonian is to express it in terms of the
density and singlet pairing operators:
\begin{eqnarray}
\Ham &=& -t \sum_{i,\alpha} [c^{\dagger}_{\alpha,i} c_{\alpha,i+1} + \text{h.c.}]
- \sum_i \mu_i \, n_i \nonumber\\
&+& \frac{U}{2} \sum_i n_i^2 + V \sum_i P_{i}^{\dagger} P_{i} \;,
\label{eq:sun-hubbardS}
\end{eqnarray}
with $U = 2 U_2$ and $V = U_0 - U_2$. This model has an exact SO(5)
symmetry~\cite{Wu2003,Wu2006}, and, for the fine-tuning $U_0 = U_2$
(or $V=0$), a SU(4) symmetry. In the latter case, the Hamiltonian
reduces to a Hubbard-like Hamiltonian with only on-site
density-density interactions. It resembles the usual SU(2) Hubbard
model, but with four colors instead of two and we refer to it in the
following as the SU(4) line. Similarly, we refer to the $U=0$ and
$V<0$ line as the BCS line since the singlet pairing is naturally
favored in this regime. The model (\ref{eq:sun-hubbardS}) has
essentially three physical parameters: the density of particles $n$,
and the two interactions $U/t$ and $V/t$ in units of the hopping $t$
(set to one in the following). Experimentally, the interacting
parameters can be varied by tuning the scattering lengths (for
instance to negative values) and the depth of the optical lattice~\cite{Bloch2007}.

In the homogeneous situation (i.e. in absence of the harmonic trap),
the phase diagram of model (\ref{eq:sun-hubbardS}) at zero temperature
has been investigated by means of low-energy approaches
\cite{Wu2005,Lecheminant2005,Lecheminant2008} and numerical
calculations \cite{Capponi2008,Capponi2007} such as the density-matrix
renormalization group (DMRG) technique
\cite{White1992,White1993,Schollwock2005} and Quantum Monte-Carlo
(QMC) simulations~\cite{Sandvik1999,Alet2005,Assaad2002}. Away from
half-filling, there are two very different spin-gapped phases which
are separated by an Ising quantum phase transition.  In the first one,
for instance along the SU(4) line with $U <0$, the BCS singlet-pairing
instability is suppressed.  The leading instability is an
atomic-density wave (ADW) with wave-vector $2k_F$ ($k_F$ being the
Fermi wave-vector) or a quartetting one.  In particular, at sufficiently
low-density, a dominant MS instability emerges which marks the onset
of the quartetting phase.  In the second spin-gapped phase, basically
obtained along the BCS line, the $2k_F$-ADW instability has now a
short-range behavior and BCS singlet pairing competes with a molecular
density-wave (MDW) with a $4k_F$ wave-vector.

In this paper, we give more details on the large-scale numerical
calculations which have been used in the short papers
\cite{Capponi2008,Capponi2007} and bring several new results. In this
respect, we present the phase diagram of model (\ref{eq:sun-hubbardS})
in absence of the trap for a generic filling which is not one atom per
site as in Ref.~\cite{Capponi2007}. Moreover, the Friedel oscillations
and excitation spectra are studied, together with the quantum phase
transition between the two spin-gapped phases from the behavior of the
entanglement entropy. We also introduce a simple observable, the
molecules fraction, which could be useful for experiments. Then, we
investigate the inhomogeneous situation and the effet of a harmonic
confining potential on the quartetting phase in order to make contact
with future experiments in spinor fermion ultracold gases. Finally,
the nature of the flux quantization in the BSC and MS phases is
analyzed in a ring geometry.

The paper is organized as follows. In
Section~\ref{sec:methods}, we recall the main results obtained
within the low-energy approach and we describe the technical details
of the three numerical methods used in this work.
Section~\ref{sec:phase-diagram} presents our main results concerning
the phase diagram and the physical properties of the phases of model
(\ref{eq:sun-hubbardS}) in the homogeneous situation.  The experimental
signatures of the quartetting phase are then discussed in
Section~\ref{sec:experiments} which includes, in particular, the
effect of the trap.  Finally, our concluding
remarks are summarized in Section~\ref{sec:conclusion}.

\section{Low-energy and numerical approaches}
\label{sec:methods}

\subsection{Low-energy approach}
\label{sec:low-energy}

\begin{table}[t]
\centering
\begin{tabular}{|c|c|c|c|c|}
\hline\hline
& \multicolumn{4}{c|}{ Phases } \\
\cline{2-5}
& \multicolumn{2}{c|}{ ADW ($K<2$) } & \multicolumn{2}{c|}{ BCS ($K>1/2$) }\\
& \multicolumn{2}{c|}{ MS  ($K>2$) } & \multicolumn{2}{c|}{ MDW ($K<1/2$) }\\
\hline
Correlator     & exponent & wave-vector & exponent & wave-vector \\
\hline\hline
$Q(x)$       & $2/K$   & 0           & $2/K$   & 0           \\
\hline
$P(x)$       & exp.     & 0           & $1/(2K)$    & 0      \\
\hline
$N_{2k_F}(x)$ & $K/2$    & $2k_F$      & exp.    & $2k_F$     \\
\hline
$N_{4k_F}(x)$ & $2K$     & $4k_F$      & $2K$     & $4k_F$    \\
\hline\hline
\end{tabular}
\caption{The possible phases, obtained by means of the low-energy
 approach, of spin-3/2 cold atoms with attractive interactions; the
 symbol exp. denotes a correlation with an exponential decay and
 the other correlations have a power-law behavior; $N_{2k_F}(x)$
 (respectively $N_{4k_F}(x)$) corresponds to the $2k_F$
 (respectively $4k_F$) part of the density correlation function.}
\label{tab:phases}
\end{table}

In this section, we recall the main results of the low-energy
approach~\cite{Wu2005,Lecheminant2005,Lecheminant2008,Controzzi2006}
on the behavior of the different order parameters that identify the
possible phases of model (\ref{eq:sun-hubbardS}). For a generic
density, the low-energy Hamiltonian separates into two commuting
pieces: a density and (hyperfine) spin part. This result is nothing
but the famous ``spin-charge'' separation which is the hallmark of 1D
incommensurate electronic systems \cite{Gogolin1998,Giamarchi2004}.
The U(1) density fluctuations remain gapless while the
spin part is fully gapped for parameters with either $U$ or $V$
negative. In the terminology of 1D electronic systems, we have the
stabilization of a Luther-Emery liquid phase
\cite{Gogolin1998,Giamarchi2004}. However, in contrast with the
standard $F=1/2$ (i.e. two-component) fermions, two very different
Luther-Emery liquid phases emerge here, which are separated by an Ising
quantum phase transition when $U =V <0$ in the weak-coupling limit.

In the first one (dubbed ADW/MS region), a representative being the
SU(4) line with $U <0$, the BCS singlet pairing with order parameter
$P_i$ displays a short-range behavior. The leading instability of this
region corresponds to the order parameter which has the slowest
(power-law) decaying correlations at zero temperature. The natural
candidates in this first phase are the density correlation $N(x) =
\langle n_i n_{i+x} \rangle$ and the quartet correlation $Q(x) =
\langle Q_i Q^{\dag}_{i+x} \rangle$ with $Q_i = c_{3/2,i} c_{-3/2,i}
c_{1/2,i} c_{-1/2,i}$.  The latter instability is rather natural since
when $U<0$ and $V=0$, the density-density term in
Eq.~(\ref{eq:sun-hubbardS}) has a tendency to form on-site molecules
(the quartets) made of four particles in all different hyperfine states.
The leading asymptotic behavior of these correlations can be computed within
the low-energy approach and the results are summarized in
Table~\ref{tab:phases}. The power-law decay of these correlation
functions depends only on the Luttinger parameter $K$ which is a
non-universal function of the interaction parameters and the density $n$.
A perturbative estimate, valid in the weak-coupling regime
$|V|/t,|U|/t \ll 1$, gives
\begin{equation}
\label{eq:pertK}
K  = 1/\sqrt{1 + [V+3U]/\pi v_F}\;,
\end{equation}
with the Fermi velocity $v_F = 2t\sin(\pi n/4)$.  Within the same
approximation, the sound velocity of the gapless density mode reads:
\begin{equation}
\label{eq:pertu}
u = v_F\sqrt{1 + [V+3U]/\pi v_F}\;.
\end{equation}
For non-interacting fermions, we have $K=1$ and $u=v_F$. From
Eq.~(\ref{eq:pertK}), we see that attractive interactions increase the
Luttinger parameter and that an artificial divergence, which occurs
when the denominator vanishes, signals the breakdown of the
perturbative calculation. From Table~\ref{tab:phases}, we deduce that
the leading instability is the $2k_F (= \pi n/2)$ ADW one for $K < 2$,
whereas the quartetting MS phase is stabilized when $K > 2$. In the
latter regime, we have an exotic Luther-Emery liquid with a
confinement of pairs (that would be objects with charge $2e$ in a context of charged particles) and the emergence of
quartets (similarly, objects with a $4e$ charge). A related Luther-Emery phase has been found
in a totally different context corresponding to the formation of multi-magnon
bound-states in the spin-1/2 $J_1$-$J_2$ Heisenberg chain under magnetic 
field~\cite{Kecke2007}. Inside the ADW/MS region, there is no
sharp quantum phase transition and only a smooth crossover. In this
respect, it might be interesting to observe that there is a continuity
between weak and strong coupling regimes in this region. Indeed, the
higher-harmonics in the quartet correlation can be estimated by means
of the low-energy approach:
\begin{equation}
\begin{split}
Q(x) \sim & A \; x^{-2/K} + B \cos(2k_F x) \; x^{-(2/K +K/2)} \\
          & + C \cos(4k_F x) \; x^{-2( K+ 1/K)} ,
\end{split}
\label{quartetcorrelboso}
\end{equation}
$A,B,C$ being non-universal amplitudes. On the other hand, along the
SU(4) line at small densities and strong attractive $U$, the physics is
essentially governed by hard-core bosons $b_i \sim Q_i$ with repulsive
interactions. The bosonic correlation function of this model is known
from the harmonic-fluid approach \cite{Haldane1981,Cazalilla2004}:
\begin{equation}
\begin{split}
 \langle b_i b^{\dag}_{i+x} \rangle 
 \sim & A \; x^{-1/2K_b} + B \cos(2 \pi \rho_0 x) \; x^{-(1/2 K_b + 2 K_b)} \\
          & + C \cos(4 \pi \rho_0 x) \; x^{-( 8 K_b + 1/2 K_b)} ,
\end{split}
\label{bosoncorrelboso}
\end{equation}
where $\rho_0 \simeq n/4$ is the density of the bosons and $K_b$ is
the underlying Luttinger parameter. From
Eqs.~(\ref{quartetcorrelboso}) and (\ref{bosoncorrelboso}), we thus
observe that there is a continuity between weak and strong coupling
regimes with $K = 4 K_b$. In particular, we also deduce an upper bound
for the Luttinger parameter $K$: $K_{\rm max} = 4$ since the value
$K_b = 1$ for non-interacting hard-core
bosons~\cite{Haldane1981,Cazalilla2004} is expected in the limit of
vanishing densities.

In the second spin-gapped region (called in the following BCS/MDW
region), obtained for instance along the BCS line, the pairing term in
Eq. (\ref{eq:sun-hubbardS}) stabilizes the order parameter $P_i$ of
the Cooper pairs. Now, the $2k_F$ ADW instability is a strongly fluctuating
order since the $2k_F$ part of the density correlation has an
exponential decay. As seen in Table~\ref{tab:phases}, the competing
orders in this phase are the BCS instability with equal-time
correlations $P(x) = \langle P_i P^{\dag}_{i+x} \rangle$ and $4k_F
(=\pi n)$ ADW operator. A BCS phase is stabilized for $K > 1/2$ which
is analogue to the standard Luther-Emery phase of spin-1/2 electrons
\cite{Gogolin1998,Giamarchi2004}. For $K < 1/2$, a MDW phase, which is
characterized by a $4k_F$ oscillation of the density fluctuations, is
predicted to emerge. For a generic filling, we expect no quantum phase
transition between BCS and MDW phases but a smooth crossover. For the 
commensurate filling of one atom per site ($n=1$), we have shown in
Ref.~\cite{Capponi2007} that a Mott transition occurs and that the MDW
phase is replaced by a Mott-insulating phase with bond ordering.

In summary, we observe that the nature of the phases found within the
low-energy approach are governed by the non-universal Luttinger
parameter $K$ which is a function of the density $n$ and the
interactions $U/t, V/t$. It is thus crucial to have a reliable
evaluation of $K$. Since model (\ref{eq:sun-hubbardS}) is not
integrable in the generic case, numerical calculations of this
parameter are required.

\subsection{Numerical methods}
\label{sec:numerics}

We use three different numerical methods to investigate the phase diagram of
model~(\ref{eq:sun-hubbardS}): mainly the density-matrix
renormalization group (DMRG), but also the exact diagonalization (ED)
and quantum Monte-Carlo (QMC) techniques.

DMRG calculations were performed at zero temperature with open
boundary conditions (OBC) using an \emph{exact} mapping of model
(\ref{eq:sun-hubbardS}) onto a two-leg SU(2) Hubbard ladder model with
special couplings (here, the spin index can take only two values
$\sigma = \pm 1/2$):
\begin{eqnarray}
  \nonumber
  \Ham^{\ladder} &=& -t_{\parallel}^{\ladder} \sum_{i,\beta,\sigma}[
  c^{\dag}_{i+1,\beta,\sigma} c_{i,\beta,\sigma} + \text{h.c.}] + U^{\ladder}
  \sum_{i,\beta} n_{i,\beta,\uparrow}
  n_{i,\beta,\downarrow} \\
\label{eq:ladder-model}
&&+ V_{\perp}^{\ladder} \sum_i n_{i,1} n_{i,2} + J_{\perp}^{\ladder}
\sum_i \mathbf{S}_{i,1} \mathbf{S}_{i,2}\;.
\end{eqnarray}
We use $\ladder$ as the label for the ladder couplings, and $\perp$ for
couplings between the two chains and $\parallel$ for couplings along
the chains. $\beta=1,2$ is the chain index, $n_{i,\beta,\sigma} =
c^{\dag}_{i,\beta,\sigma} c_{i,\beta,\sigma}$, $\mathbf{S}_{i,\beta}$
is the spin operator, and $n_{i,\beta} = \sum_{\sigma}
n_{i,\beta,\sigma}$ the local density. For the hoppings, we have
$t_{\parallel}^{\ladder} = t$ and $t_{\perp}^{\ladder} = 0$. For the
on-site interaction, $U^{\ladder} = U$. For the next-nearest neighbor
density-density interaction on rungs, $V_{\perp}^{\ladder} = U + V/2$,
and also a Heisenberg coupling on the rungs $J_{\perp}^{\ladder} =
-2V$. All other couplings are equal to zero. In this mapping,
the local number of states per site is strongly reduced as it is
$2^2=4$ for a SU(2) Hubbard site and $2^4=16$ for a spin-3/2 Hubbard
site. Symmetries are used to fix the total number of fermions to $N_f$
and the total $z$-component of the spin to zero. We have typically
kept 1000 states of the reduced density matrix, and sometimes up to
1400. Convergence depends on the regions of the phase diagram, and is
harder with the trap. On the SU(4) line with large $\vert{U}\vert$, the convergence
is very good as the physics is
essentially the one of hard-core bosons. The discarded weight
typically ranges from $10^{-11}$--$10^{-8}$ when both interactions are
negative (and not to small) to $10^{-6}$ if one is positive or small.
Moreover, the discarded weight decreases with density so that
simulations become easier and more accurate in this regime.

On the SU(4) line with total $S^z=0$, the numbers of particles
$N_{\sigma}$ per specie are independently conserved. Therefore, with an
appropriate choice of boundary conditions and $N_{\sigma}$ (for instance
periodic boundary conditions and $N_{\sigma}$ odd), the particles do
not experience any statistics so that, by means of a Jordan-Wigner
transformation, the model is \emph{strictly equivalent} to a hard-core
boson model on a \emph{four}-leg ladder for which chains are only coupled via a
density-density interaction term $V_{\perp}^{\ladder} = U$ between all
chains. Such a bosonic model has no sign problem and can be
efficiently simulated by QMC techniques such as the Stochastic Series
Expansion (SSE) algorithm~\cite{Sandvik1999,Alet2005}. We use the ALPS
software implementation of the SSE
algorithm~\cite{Albuquerque2007,Troyer1998}. Note that, contrarily to
DMRG, the algorithm works in the grand-canonical ensemble and at
finite temperature. Away from the SU(4) line, Fermi statistics cannot
be avoided and therefore, we have also used a determinantal QMC
algorithm (DQMC), which has no sign problem over a relatively wide
range of parameters~\cite{Capponi2007}. For this algorithm, we have
used the projector approach that provides ground-state properties with
a fixed number of particles~\cite{Assaad2002}.

\section{Phase diagram}
\label{sec:phase-diagram}

This section gathers results on the phase diagram for a generic
density, i.e. a density for which no commensurability effects are
expected, and which is sufficiently low to realize the MS phase on the
SU(4) line~\cite{Capponi2008}. We choose $n=0.75$. We first explain how the Luttinger
parameter $K$ is computed numerically, before giving more details on
the physics of each phase.

\subsection{Extracting the Luttinger exponent}

\begin{figure}[b]
\centering
\includegraphics[width=0.95\columnwidth,clip]{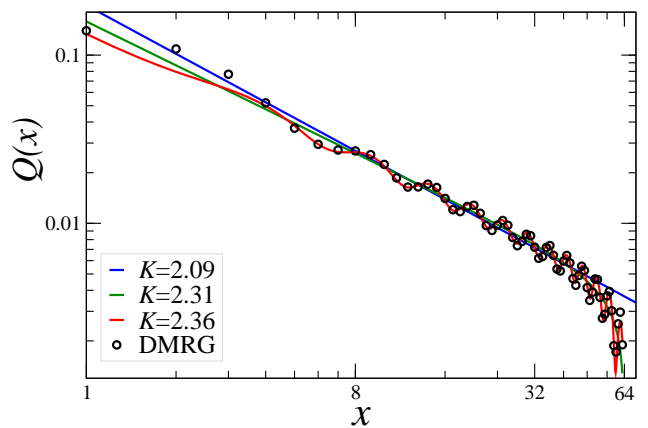}
\caption{Typical fit of the quartet correlations in the MS phase on
  the SU(4) line. Data are obtained by DMRG with $L=128$ and $U/t=-4$
  at filling $n = 0.75$. See text for the three fitting functions.
  Using Eq.~(\ref{eq:fitcorr}) gives the Luttinger parameter $K =
  2.36$.}
\label{fig:fitcorr}
\end{figure}

As one can see from Table~\ref{tab:phases}, the Luttinger exponent $K$
can be extracted from algebraically decaying correlation functions.
For instance, the quartet (or MS) correlations $Q(x)$ gives access to
$2/K$ in all regions of the phase diagram. They can be reliably
computed with DMRG if the number of state kept is sufficiently large.
As data are computed on finite and open chains, there is no
translational invariance and all correlators depend on both
positions of the sites. For instance, $Q(x) = \langle Q_i
Q^{\dag}_{i+x} \rangle$ will actually depend on $i$. We fix $i=m=L/2$
to be at the middle of the chain and control finite size effects using
results from conformal theory~\cite{Cazalilla2004}. By denoting the
conformal distance $d(x\vert L)=L \vert \sin(\pi x/L)\vert /\pi$, the
leading term of the quartet correlations is of the bosonic form
\begin{equation}
\label{eq:fitcorr}
\begin{split}
Q(x) =& \rho_0 \sqrt{1+ c_0 \frac{\cos(\pi n x/2 +\delta)}{[d(2(m-x)\vert 2L)]^{K/4}}}\\ 
& \qquad\times \left[\frac{\sqrt{d(2x\vert 2L) d(2m\vert2L)}}{d(x+m\vert 2L)d(x-m\vert 2L)} \right]^{2/K} \;.
\end{split}
\end{equation}
If one writes the $Q_i$ operator in a density-phase representation
$\sqrt{q_i} e^{i\phi(x_i)}$, the first term is $\sqrt{q_i q_{i+x}}$
where $q_i = \langle Q^{\dag}_i Q_i \rangle$ denotes the local density
of quartets (bosons). Because of OBC, Friedel oscillations appear
close to the edge, leading to a typical $2k_F$ cosine term that decays
algebraically \emph{from} the edge (see a discussion in
Sec.~\ref{sec:friedel}). In terms of bosons, this decay of the density
fluctuations~\cite{Cazalilla2004} is controlled by $K_b$ which gives
$K/4$ for quartets. Note that the scaling dimension of the density
operator is also $K/4$. Thus, our fitting procedure stems more from the
phenomenology of hard-core bosons than from an exact result. Note that oscillations cannot
be explained by the harmonics of the quartetting correlations as
derived in Eq.~(\ref{quartetcorrelboso}) because the exponents of the
sub-leading terms are of order 2 and 5, which are far too
large to explain the oscillations (which actually increase with $x$).
The amplitude $c_0$ and phase-shift $\delta$ are unknown parameters. The second
term is the leading algebraic decaying term $x^{-2/K}$ modified by
finite size effects. The function replacing $1/x$ in between the
brackets accounts for the vanishing of the wave-function at the edge
of the box which induces a drop in the correlations. Typical data in
the MS phase are given in Fig.~\ref{fig:fitcorr} for a chain with size
$L=128$. Rather strong Friedel oscillations are observed in the signal
(in the BCS phase, these oscillations are much smaller). Three fits
are used to extract $K$. Firstly, a simple algebraic fit (which
corresponds to taking $c_0=0$ and $L=\infty$ in
Eq.~(\ref{eq:fitcorr})) yields $K = 2.09$. Secondly, a fit without the
$2k_F$ oscillations ($c_0=0$) gives $K=2.31$. Thirdly, a fit using
Eq.~(\ref{eq:fitcorr}) with $\rho_0$, $c_0$, $\delta$ and $K$ as free
parameters gives $K=2.36$ and an excellent agreement with the data. It
is thus important to take into account the finite size effects to have
a reliable evaluation of $K$. In a previous work~\cite{Capponi2007},
we have used an averaging of the correlators over $i$; this suppresses
the oscillations but gives a less accurate estimate for $K$. A similar fit
function as in Eq.~(\ref{eq:fitcorr}) but with a phenomenologically
introduced cosine oscillations was used in Ref.~\cite{Capponi2008} and leads to results
very close to the ones obtained from Eq.~(\ref{eq:fitcorr}).

\begin{figure}[t]
\centering
\includegraphics[width=0.85\columnwidth,clip]{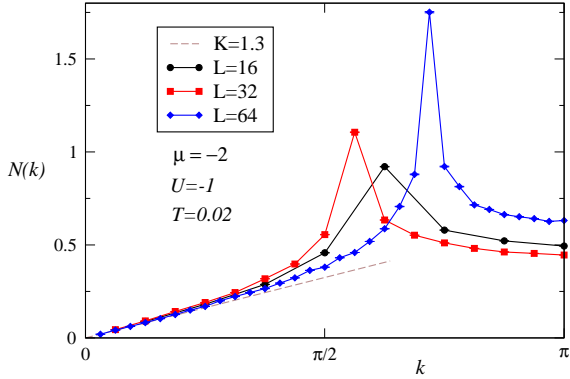}
\caption{Density structure factor $N(k)$ obtained from QMC SSE
  simulations at low temperature ($T=0.02 t$), fixed chemical
  potential $\mu=-2t$ and for various sizes $L$. Since finite-size
  effects are rather small, an accurate estimate of $K$ can already be
  obtained on small systems from the small-$k$ linear behavior. Here,
  for $U/t=-1$ and a density close from $1.2$, one gets $K\simeq 1.3$
  which is compatible with the DMRG estimate.}
\label{fig:nkfit}
\end{figure}

Another systematic way of calculating the Luttinger exponent is to use
the density correlations $N(x)$ and the associated structure factor $N(k)$,
where $k$ is the wave-vector. This method is particularly suited for
QMC as the density operator can be more easily sampled than the
quartet operator. The value of $K$ is extracted from the small
wave-vector behavior of $N(k)$:
\begin{equation}
K=\frac{2\pi}{4} \lim_{k \rightarrow 0} \frac{N(k)}{k},
\label{eq:KfromQMC}
\end{equation}
with a factor 4 in the denominator corresponding to the number of fermionic flavors. For
instance, this procedure has been shown to be very accurate for the
spin-1/2 Hubbard model~\cite{Ejima2005}. An example of a typical fit
is given on Fig.~\ref{fig:nkfit} at fixed chemical potential $\mu$.
Note that, since the QMC SSE algorithm is grand canonical, the density
will slightly vary when the parameters (temperature or size) are
changed\footnote{Note that canonical algorithms are also available for
  such models.}. This effect can be seen from the position of the
$2k_F=\pi n /2$ peak in Fig.~\ref{fig:nkfit}. One advantage is that a
linear fit is simple to perform. However, in the limit of small
densities, the $2k_F$ peak approaches 0 which makes it difficult to
find the linear small-$k$ regime on finite size systems.

\subsection{Phase diagram at the generic density $n=0.75$}

\begin{figure}[t]
\centering
\includegraphics[width=0.95\columnwidth,clip]{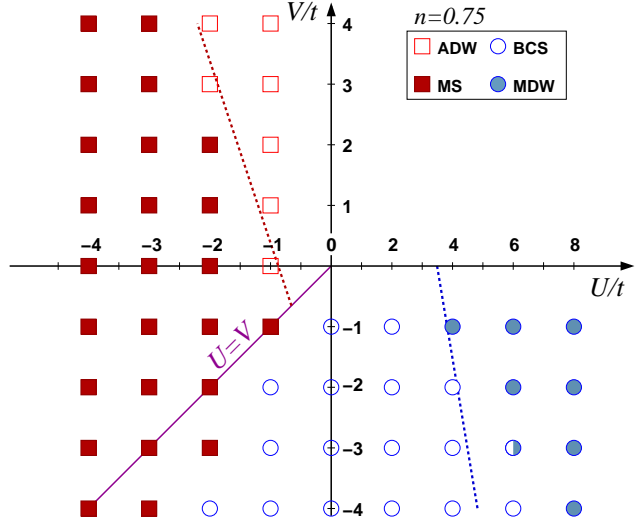}
\caption{Phase diagram of the spin-3/2 Hubbard model
  (\ref{eq:sun-hubbardS}) at the incommensurate filling $n=0.75$, and
  for attractive interactions ($U\leq 0$ or $V\leq 0$) from DMRG
  calculations (see text for definitions of the phases). The $U=V$
  line is the perturbative estimate for the transition between the two
  regions BCS/MDW and the ADW/MS regime. The dashed lines are the
  perturbative estimates for the crossovers between respectively
  BCS/MDW and ADW/MS.}
\label{fig:phase-diagram}
\end{figure}

Fig.~\ref{fig:phase-diagram} displays the phase diagram of the
spin-3/2 Hubbard model for attractive interactions ($U\leq 0$ or
$V\leq 0$) and a density $n=0.75$. The density is chosen in such a way
that the MS exists (from Ref.~\cite{Capponi2008} we know that this is
the case on the SU(4) line), and that there are no commensurate
phases\footnote{Actually, one could argue that $n=3/4$ is a simple
  fraction and that commensurate phases can occur. However, in terms
  of bosons, this would correspond to a density $3/16$ for which a
  Luttinger exponent $K_b=2/16^2$ is required to drive the
  transition~\cite{Giamarchi2004}, giving $K=1/32$ which is very
  small, but could, in principle, still appear at very large
  interactions.} (in contrast to $n=1$
\cite{Lecheminant2005,Capponi2007} or $n=2$). Note that the MS phase
is not accessible at filling $n=1$ while the
perturbative estimate of Eq.~(\ref{eq:pertK}) predicts its existence \cite{Capponi2007}.
From its wide extension in Fig.~\ref{fig:phase-diagram}, we observe
that the MS phase, is very robust under the symmetry breaking term
$V$. Thus, the quartet molecular phase is not an artifact of the SU(4)
symmetry. This is an important result since in most of the realistic
situations, the actual symmetry is expected to be smaller than SU(4).
Part of the answer is given in 1D systems by the accepted view that,
at sufficiently low energies and for generic interactions, the
dynamical symmetry is most likely to be enlarged~\cite{Lin1998}:
though the SU(4) symmetry is not an exact symmetry, it is physically
meaningful as an effective low-energy theory. As a consequence, the
SU(4) model studied in Ref.~\cite{Capponi2008} is a very good starting
point to explore the main features of the quartet phase.

As a remark, we argue that the quartet phase also emerges in a problem
with no extended SO(5) symmetry. For instance, we can consider the
two-leg ladder model described in Sec.~\ref{sec:numerics} on the SU(4)
line.  The model simplifies to a model of two spin-1/2 Hubbard chains
coupled only by the inter-chain density-density interaction
$V^{\ladder}_{\perp}$. If we relax the constraint $V^{\ladder}_{\perp}
= U$ and let $V_{\perp}$ vary, we have the following picture. For
$V_{\perp}=0$, the SU(2) Hubbard chains are exactly solvable by the
Bethe-ansatz technique and the corresponding Luttinger parameters
$u_{\text{SU}(2)}, K_{\text{SU}(2)}$ can be computed
exactly~\cite{Giamarchi1995}. Bosonizing the $V_{\perp}$ coupling
between the two chains gives, for the symmetric combination of the
modes, the Luttinger parameter
\begin{equation}
\label{eq:sun-ladder-mapping}
K_{+} = K_{\text{SU}(2)} / \sqrt{1 + V_{\perp} K_{\text{SU}(2)}/\pi u_{\text{SU}(2)}}\,.
\end{equation}
$K_+$ can be identified with $K$ when $V_{\perp} = U$ and governs the
quartetting correlations as in Eq.~(\ref{quartetcorrelboso}). We see
that a negative $V_{\perp}$ gives $K_+ > K_{\text{SU}(2)}$. Yet, we
know from Ref.~\cite{Giamarchi1995} that $K_{\text{SU}(2)} \rightarrow
2$ in the limit of low densities and negative $U_{\text{SU}(2)}$. A
finite negative $V_\perp$ should thus easily stabilize a superfluid
quartet phase with $K_{+}>2$. Furthermore, we must note that
Eq.~(\ref{eq:sun-ladder-mapping}) is perturbative in $V_{\perp}$ but
otherwise valid for \emph{arbitrary} values of $u_{\text{SU}(2)}$ and
$K_{\text{SU}(2)}$, which are known even in the strong coupling regime
$|U_{\text{SU}(2)}|/t \gg 1$. In particular, the fact that $K$
saturates along the $U/t$ line of the SU(4) Hubbard
model~\cite{Capponi2008} might be similar to the saturation of
$K_{\text{SU}(2)}$ at large $\vert{U_{\text{SU}(2)}}\vert$. For both
models, the saturation is associated with the onset of a hard-core
boson regime of pairs for SU(2), and quartets for SU(4). A similar
superfluid phase where pairing of bosons along the rungs occurs in the
bosonic ladder model~\cite{Orignac1998}, the phenomenology and
perturbative argument being essentially the same.

\subsection{Excitations gaps and spectra}
\label{sec:gaps}

We now turn to the excitation gaps in the two regions of the phase
diagram. In the context of cold atoms experiments, the quartet phase
can be probed by radio-frequency spectroscopy to measure the
excitation gaps of the successive quartet dissociation process. In
particular, the existence of molecules can be characterized by finite
one and two particle gaps while four particle excitations remain
gapless. This feature alone does not distinguish between various
phases (dominant superfluid or density correlations) but it allows to
check for the formation of bound-states.

\begin{figure}[t]
\centering
\includegraphics[width=0.95\columnwidth,clip]{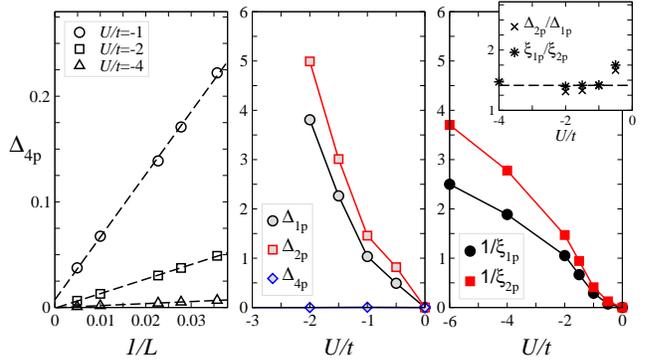}
\caption{One, two and four particle gaps along the SU(4) line for the
  density $n=1$. \emph{Left:} scaling of the four particles gap for different $U$. \emph{Middle:} $\Delta_{1\text{p}}$ and $\Delta_{2\text{p}}$ as a function of $U/t$ on a finite system with $L=12$, and extrapolated $\Delta_{4\text{p}}$. \emph{Right:} We also show the inverse of the correlation lengths
  $\xi_{1\text{p},2\text{p}}$ obtained from the Green's function and the pairing
  correlations. \emph{Inset:} comparison of the numerically obtained ratios and the $\sqrt{2}$ prediction.}
\label{fig:gaps}
\end{figure}

The energy gap to fill when adding $p$ particles in the
system is defined by
\begin{equation}
\label{eq:gaps}
\Delta_{p\text{p}} = E_0(N_f+p) + E_0(N_f+p) - 2E_0(N_f)\;,
\end{equation}
with $E_0(N_f)$ the energy of the ground-state with $N_f$ particles.
We choose $N_f=4(2m+1)$, with $m$ an integer, so that we would have
closed shells in the case of periodic boundary conditions.
Fig.~\ref{fig:gaps} provides the results on the one, two and four
particles gaps on the SU(4) line with a density $n=1$ and a small
system $L=12$. For large sizes and interaction $\vert U\vert$,
quartets are strongly bound and convergence of systems with $N_f$ not
a multiple of four can fail. Indeed, even after many sweeps, the
density distributions are not symmetrical with respect to the center
of the chain. Results given in Fig.~\ref{fig:gaps} are those with
symmetrical ground-states and well-converged energies. The small size
of the chain may cause finite size effects but, still, a clear opening
of the one and two particle gaps is found. For $\Delta_{4\text{p}}$,
no convergence issues are found and scaling can be performed.
Fig.~\ref{fig:gaps} shows that $\Delta_{4\text{p}}=0$ in the
thermodynamical limit, in agreement with the algebraic decay of the
correlations. To further check the consistency of the numerics and the
low-energy theory, we use the fact that the ratio $\Delta_{2\text{p}}
/ \Delta_{1\text{p}}$ is known to be exactly $\sqrt{2}$ from the
integrability of the SU(4) Thirring model~\cite{Andrei1980} describing
the spin part of the Hamiltonian in the low-energy approach. This
ratio is plotted for the $L=12$ gaps in the inset of
Fig.~\ref{fig:gaps} and agrees reasonably well with the prediction.

Another way of probing the presence of finite gaps is to look at the
associated correlation functions. The one and two particles gaps, when
finite, are associated with the exponential decay of the Green's and
pairing correlation functions, as observed in Ref.~\cite{Capponi2007}.
In this case, the two correlation lengths behave as $\xi_{1\text{p},2\text{p}}
\sim u/\Delta_{1\text{p},2\text{p}}$, where $u$ is the sound velocity of the
bosonic mode. The inverse correlation lengths are given as a function
of interaction in Fig.~\ref{fig:gaps}. The same universal ratio is
expected for the correlation length and well observed numerically on
Fig.~\ref{fig:gaps}. These results support the correctness of the
low-energy approach, even for strong couplings, and in particular the
validity of the separation of the spin-charge sectors.

\begin{figure}[t]
\centering
\includegraphics[width=0.95\columnwidth,clip]{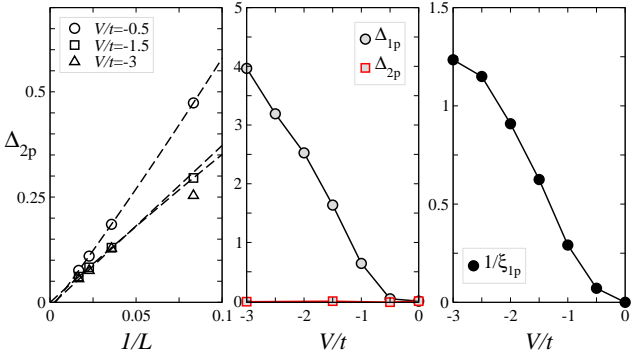}
\caption{One and two particle gaps along the BCS line for the density
  $n=1$. \emph{Left:} Scaling of $\Delta_{2\text{p}}$. \emph{Middle:} $\Delta_{1\text{p}}$ on a finite system with $L=12$, and extrapolated $\Delta_{2\text{p}}$. \emph{Right:} the inverse correlation length of the Green's function
  $\xi_{1\text{p}}$ is also shown.}
\label{fig:BCSgaps}
\end{figure}

Similarly, gaps and the inverse correlation length can be computed by
DMRG on the BCS line. Results are given in Fig.~\ref{fig:BCSgaps}
which shows a smooth opening as soon as interactions are turned on.
Two and four particle excitations are gapless while a one-particle gap
opens. We only show $\Delta_{2\text{p}}$ because $\Delta_{4\text{p}}$
also scales to zero when $\Delta_{2\text{p}}=0$ in the thermodynamical
limit. These behaviors follow the results obtained in
Ref.~\cite{Capponi2007} for the pairing and correlation functions. To
give additional insights on the opening of the one-particle gap, we
provide the evolution of the inverse correlation length of the Green's
function $\xi_{1\text{p}}$. Experimentally, the one-particle
correlation length $\xi_{1\text{p}}$ will appear in the momentum
distribution of the condensate.

\begin{figure}[t]
\centering
\includegraphics[width=\columnwidth,clip]{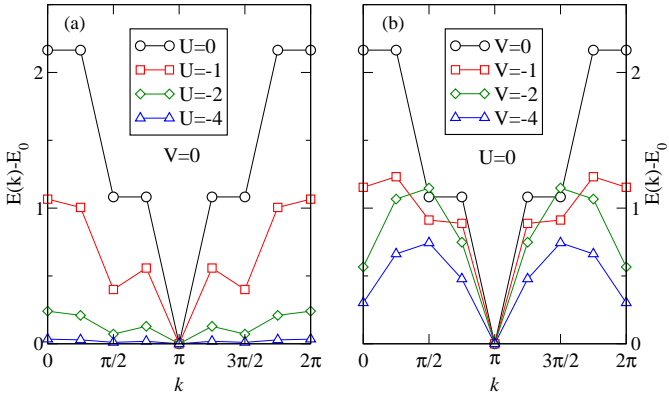}
\caption{Lowest energy $E(k)$ as a function of the momentum $k$ for a
  ring of length $L=8$ with 8 fermions for various interactions (a)
  along the SU(4) line ($V=0$); (b) along the BCS line ($U=0$).
  Antiperiodic boundary conditions are chosen to have closed shells.
  Energies are measured relative to the ground-state energy.}
\label{fig:Evsk}
\end{figure}

Exact diagonalization on small systems on a ring allows for the
computation of the excitation energy spectrum $E(k)$ vs. momentum $k$.
Even if finite size effects can be important, some qualitative
information can be extracted. Figure~\ref{fig:Evsk} displays the
spectra along the SU(4) and BCS lines for the density $n=1$. Note that
anti-periodic boundary conditions are used to have closed shells when
$U=V=0$ for a chain with $4(2m)$ fermions with $m$ an integer. On the
SU(4) line, we observe that the $2k_F$ excitation has a lower energy
than the $4k_F$, which is associated with the dominant density
fluctuations at $2k_F$ in this region of parameters. As
$\vert{U}\vert$ increases, all energies go down, so the sound
velocity of the charge mode $u$ also decreases, in agreement with the
perturbative estimate of Eq.~(\ref{eq:pertu}). The spectrum evolves
continuously towards the strong-coupling limit. On the contrary, along
the BCS line, a crossover is found between a regime, at low
$\vert{V}\vert$, in which the minimum is at $2k_F$, and the strong
coupling regime for which the minimum is at $4k_F$. We will see
hereafter that a similar crossover is found in the density
fluctuations and Friedel oscillations. Lastly, one can note that the
sound velocity $u$ slowly decreases through this crossover line (and
slower than on the SU(4) line), again in agreement with the
perturbative prediction.

\subsection{Quartets formation on the SU(4) line}

\begin{figure}[t]
\centering
\includegraphics[width=0.75\columnwidth,clip]{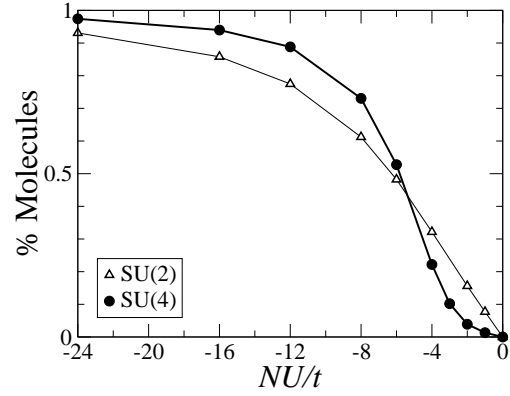}
\caption{Fraction of molecules in the system as a function of $U$ for
  the SU(2) and SU(4) models as defined by Eq.~(\ref{eq:fraction})
  from a system with $L=64$ and density $n=1$.}
\label{fig:quartetfraction}
\end{figure}

In this section, we discuss the crossover from the weak-coupling
regime to the strong-coupling regime on the SU(4) line as the
attractive interaction is increased. When $\vert{U}\vert$ is large,
the physics is essentially the one of hard-core bosons with repulsive
interactions, as it has been discussed in Sec.~\ref{sec:low-energy}
within the low-energy approach. To investigate how the quartets form,
we can compute the local density of these ``molecules''. From a more
general point of view, and to compare with the SU(2) case, we consider
$N$-particle bound states in the context of the SU($N$) Hubbard
model~\cite{Capponi2008}. The local density of molecules is $m(x) =
\langle{n_{x,1} \cdots n_{x,N}}\rangle$ (which we denote by $q(x)$ for
quartets). For free fermions, this operator has a finite expectation
value that we subtract to keep only the connected part $m(x) =
\langle{n_{x,1} \cdots n_{x,N}}\rangle - (n/N)^N$. If molecules are tightly bound on-site, we expect $\langle{
  n_{x,1} \cdots n_{x,N}}\rangle$ to be close to $n/N$ though slightly
lower. Therefore, we can define a molecule fraction (number between zero
and one) as
\begin{equation}
\label{eq:fraction}
\%\text{Molecules} = \frac{\overline{m(x)} - (n/N)^N}{n/N - (n/N)^N}\;,
\end{equation} 
where the bar means averaging over all sites. The evolution of this
quantity along the SU(2) and SU(4) lines are compared in
Fig.~\ref{fig:quartetfraction} as a function of $NU$ (and not $U$)
because the interaction term scales like $N^2$ while the kinetic term
only scales as $N$. At large $\vert{U}\vert$, molecules are tightly
bound, but the SU(4) model is closer to a hard-core boson model than
the SU(2) one. Another difference is the low-$U$ increase which is
linear for SU(2) and power-law for SU(4) with an exponent larger than
2. Note that the behavior should also depend on density, particularly
at small $U$.

Consequently, we expect the large negative $U$ physics to be
essentially the one of hard-core boson. Still, we emphasize a major
difference between SU(2) and SU(4): from perturbation theory, the
SU(2) case leads to hard-core bosons with equal effective hopping and
nearest-neighbor repulsion (also equivalent to an effective spin-1/2
XXZ chain~\cite{Xianlong2007}); on the contrary, in the SU($N$) case
(with $N>2$), the effective hopping at $N^{\text{th}}$ order in perturbation
theory behaves as $t^N/|U|^{N-1}$, so it is negligible compared to
nearest-neighbor repulsion, which is of order $t^2/|U|$.

\subsection{Evolution of the Luttinger parameter and the commensurate 
phase ADW$^{\pi}$ for $n=2$}

\begin{figure}[t]
\centering
\includegraphics[width=0.65\columnwidth,clip]{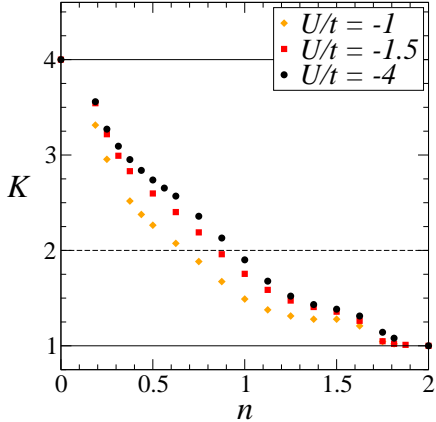}
\caption{(Color online) Luttinger parameter $K$ vs filling $n$ for
  various $U$ obtained by fitting quartet correlations as in
  Fig.~\ref{fig:fitcorr} for the SU(4) model. $K\rightarrow 4$ when
  $n\rightarrow 0$ and $K \rightarrow 1$ when $n \rightarrow 2$ are
  the asymptotic behavior of the strong-coupling limit.}
\label{fig:KSU4}
\end{figure}

The previous considerations allow for a simple interpretation of the
behavior of the Luttinger parameter $K$ as a function of the density
$n$ for large negative $U$. DMRG results are given in
Fig.~\ref{fig:KSU4}. As expected from the strong coupling argument,
$K$ decreases from $K=N$ to $K=N/4$ (from 4 to 1) as the density $n$
varies from 0 to half-filling ($n=2$). Particle-hole symmetry would give
the behavior for $2 \leq n \leq 4$. Again, we recall that molecular superfluidity
is the dominant instability when $K>2$, which is generically the case
at low enough density. When $n=2$, a fully gapped phase is obtained
with short-range quartet correlations. The phase is two-fold
degenerate with a $\pi$ ordering of the local density (one quartet
every two sites). Hence, we call this phase ADW$^{\pi}$. In terms of an
effective bosonic model discussed in the previous paragraph, this
corresponds to a ``charge'' density wave phase of the equivalent bosonic
model at half-filling~\cite{Zhao2006,Zhao2007}.  The density of bosons
being 1/2, the corresponding critical value for their Luttinger
parameter~\cite{Giamarchi2004} is $K_b=1/2$ (at fixed density,
changing interactions), which gives $K=2$ for our model. As $K=1<2$,
the ADW$^{\pi}$ phase emerges as soon as the interaction $U$ is turned
on.  Working at fixed interactions and varying the density, the
critical value is now $K_b=1/4$, which gives the observed limiting
value $K=1$ as one approaches $n=2$ (see also
Ref.~\cite{Capponi2008}).

\subsection{Density fluctuations and Friedel oscillations}
\label{sec:friedel}

The density fluctuations can be analyzed from the correlations
structure factor $N(k)$ with QMC, and from the behavior of the Friedel
oscillations of the local density in open chains, as usually done in
DMRG. The Friedel oscillations are the response of the fermionic
density to the open end of the chain, which acts as an impurity. These
modulations can give access to Luttinger parameters~\cite{Egger1995}.
Data for the SU(4) line (not shown), and more generally in the ADW/MS
region of the phase diagram, are all consistent with the low-energy
predictions~\cite{Wu2005,Lecheminant2005,Capponi2007} $N(x) \sim
\cos(2k_F x) x^{-K/2}$ for the density correlations and $n(x) \sim
\cos(2k_F x) x^{-K/4}$ for the Friedel oscillations. Note that the
$N(2k_F)$ peak diverges with the system size $L$ provided $K < 2$,
signaling the quasi-ordering of the density fluctuations of the ADW
phase~\cite{Capponi2007}.

\begin{figure}[t]
\centering
\includegraphics[width=0.95\columnwidth,clip]{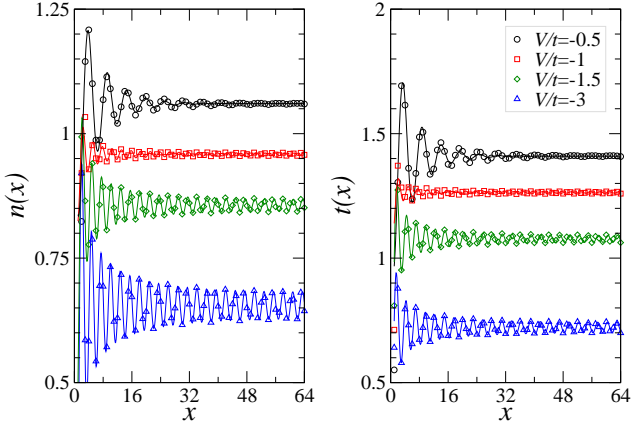}
\caption{Local density $n(x)$ and kinetic energy $t(x)$ along the BCS
  line at filling $n=0.75$ on a $L=128$ chain for increasing
  $\vert{V}\vert$. Friedel oscillations of the density have been
  shifted vertically for clarity. Thin lines are fits to results from
  DMRG calculations (see text).}
\label{fig:LocalBCS}
\end{figure}

Friedel oscillations in the BCS phase have a different behavior. As
shown in Fig.~\ref{fig:LocalBCS} for the generic density $n=0.75$,
there is a qualitative change in the wave-vector of the oscillations
from $2k_F$ at low $\vert{V}\vert$ to $4k_F$ at large $\vert{V}\vert$.
The predictions are that the $2k_F$ term should be short-range but a
$4k_F$ term can develop with correlations $N(x) \sim \cos(4k_Fx)
x^{-2K}$. For the Friedel oscillations, we thus expect a leading
contribution behaving as $n(x) \sim \cos(4k_Fx) x^{-K}$, similar to
what was found in two-leg ladders~\cite{White2002}. To explain the
behavior observed in Fig.~\ref{fig:LocalBCS}, we argue that at low
$\vert V\vert$, the amplitude of the $2k_F$ term remains significant (it is
finite for a free system at $V=0$) and with a correlation length which
is still large (see Fig.~\ref{fig:BCSgaps}). When $\vert V\vert$
increases, the $4k_F$ term emerges with an increasing amplitude. Fits
have been carried out in Fig.~\ref{fig:LocalBCS} using $n(x) = n_0 +
n_1 \cos(2k_F x +\delta) e^{-x/\xi}$ for $V/t=-0.5$ and $n(x) = n_0 +
n_1 \cos(4k_Fx +\delta)/ [d(x|L)]^K$ for larger $\vert V\vert$. The
Luttinger exponents obtained from the fits are close to the ones
obtained from the pairing correlations. In addition to DMRG
calculations, DQMC data (see for instance Fig.~4 of
Ref.~\cite{Capponi2007} and other results not shown) support a similar
qualitative change in the wave-vector and no divergence of the $4k_F$
amplitude with the system size. Indeed, this divergence only occurs
for $K<1/2$, i.e. in the MDW phase. Lastly, the same crossover around
$V/t=-1$ is found in Fig.~\ref{fig:Evsk}(b).

The crossover to the large $\vert V\vert$ physics can be qualitatively
understood within the following picture: when $V$ is large, Cooper
pairs have a tendency to form on-site, and certainly repel each other
to gain local kinetic energy. This gives a typical $4k_F$ fluctuation
of the local density and kinetic energy as found in
Fig.~\ref{fig:LocalBCS}. The local kinetic energy term is $t(x) =
\langle \sum_{\sigma} c^{\dag}_{x+1\sigma} c_{x\sigma}\rangle$. In the
ADW/MS region, it follows the Friedel oscillations of the density, so
we have $t(x) \sim \cos(2k_Fx) x^{-K/4}$. In the BCS/MDW region, the
$2k_F$ component is short-range so the leading term is the $4k_F$ one,
$t(x) \sim \cos(4k_Fx)x^{-K}$, as for $n(x)$.  The enhancement of
these fluctuations (the total kinetic energy rather decreases with
interactions as seen in Fig.~\ref{fig:LocalBCS}) as $\vert V\vert$ is
increased is reflected through the decrease of the Luttinger exponent
$K$, as it was found for $n=1$ in Fig.~5 of Ref.~\cite{Capponi2007}. A
similar slow decrease with values of $K$ lower than one at large
$\vert V\vert$ is found for $n=0.75$.  This decrease is not predicted
in the perturbative estimate of Eq.~(\ref{eq:pertK}) and is therefore
a typical strong-coupling behavior. If one adds the repulsive
interaction $U$ on-site, strictly on-site pairs are no more favored
and the pairs lower their energy by delocalizing themselves on a bond.
If $U$ is large enough and the density commensurate at $n=1$, this
qualitative picture leads to the bond-order wave phase observed in
Fig.~7 of Ref.~\cite{Capponi2007} which breaks translational symmetry
and is two-fold degenerate.

 \subsection{The transition between BCS/MDW and ADW/MS}

This section gives some results on the transition line between the two
regions BCS/MDW and ADW/MS. It was already shown that it belongs to
the Ising universality class and that the ratio between the pairing
and quartet correlations $R(x)=P^4(x)/Q(x)$ has the universal behavior
$1/x$ at the critical point in agreement with conformal field theory (CFT)
predictions~\cite{Lecheminant2005,Lecheminant2008,Capponi2007}.

\subsubsection{Effect of the density on the transition line}
\label{sec:TransitionDensity}

\begin{figure}[t]
\centering
\includegraphics[width=0.98\columnwidth,clip]{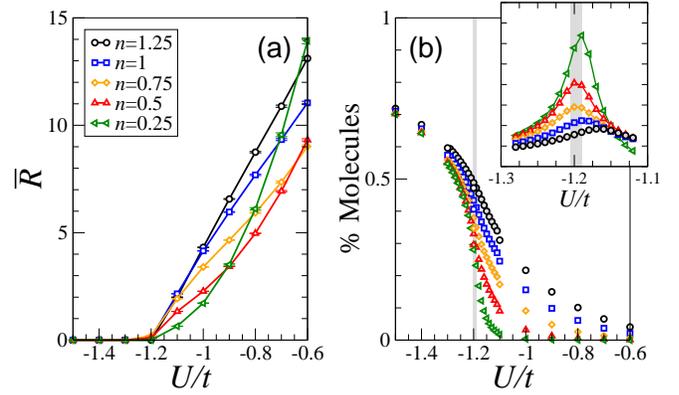}
\caption{Effect of the density on the transition line with fixed
  $V/t=-2$ (a) Averaged ratio $\overline{R}$ signaling the BCS/MS
  transition for different densities. The critical point $U/t \simeq
  -1.2$ of the BCS/MS transition hardly depends on density, even in
  the strong coupling limit. (b) The fraction of molecules (quartets)
  does not give a direct estimate of the transition but its derivative
  (inset) does. Open symbols are for a system with $L=128$ while
  symbols filled with grey correspond to $L=32$ (finite size effects
  are stronger at large density but remain small).}
\label{fig:TransitionDensity}
\end{figure}

We first investigate the question of the dependence of the transition
line with respect to the density $n$. This is an important issue for
inhomogeneous systems, such as trapped cold atoms, since the local
density in the cloud evolves continuously from zero to a finite value in the bulk.
We know that the perturbative result $U=V$ for this line does not
depend on $n$, while the crossover lines in
Fig.~\ref{fig:phase-diagram} noticeably depend on $n$ from Eq.~(\ref{eq:pertK}).
In the strong coupling regime, we study numerically the transition
line for several densities. To that purpose, we first use the behavior
of $R(x)$ averaged over distances ranging from $x=20$ to 40 (from
correlations in a system with $L=128$), which we denote by
$\overline{R}$.  In Ref.~\cite{Capponi2007}, the same ratio was used
for a fixed distance $x=45$ for $n=1$; the two procedures lead to the
same results but the first one is more suited to systems with low
densities as oscillations have a longer wave-length.  From
Fig.~\ref{fig:TransitionDensity}(a), we see that $\overline{R}$
vanishes linearly around the critical point which is due to the Ising
nature of the transition~\cite{Capponi2007}. However, the higher the
density, the wider the range of $U$ over which the linear behavior is
observed. At density $n=0.25$, the linear behavior is not recovered,
certainly because of our mesh points. The main result is that the
position of the critical point $U/t \simeq -1.2$ hardly depends on the
density, though corresponding to the strong coupling regime.

Experimentally, it would be impossible to measure $\overline{R}$ so it
is interesting to look at the behavior of the molecule fraction
through the transition. This is given in
Fig.~\ref{fig:TransitionDensity}(b) in which one finds that the
molecule fraction increases from the BCS/MDW to the ADW/Ms region, but
rather smoothly at large densities.  However, taking the derivative of
the curve with respect to $U$ gives a clear peak pointing at the
critical point. Consequently, this would be a rather simple way to
locate the transition line experimentally as it could be measured and
works even for an inhomogeneous cloud. The dependence of the peak with
respect to the density is found to be small.

\subsubsection{Entanglement entropy and central charge}

\begin{figure}[t]
\centering
\includegraphics[width=0.95\columnwidth,clip]{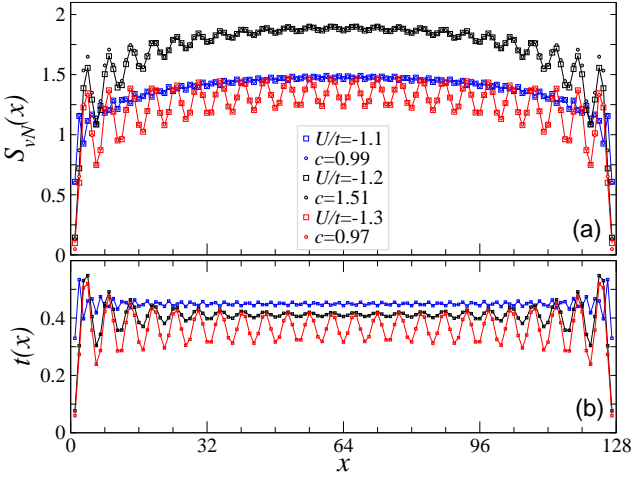}
\caption{Von Neumann block entropy $S_{\text{vN}}(x)$ for a block of
  size $x$ and local kinetic energy $t(x)$ around the critical point
  $U/t=-1.2$ for fixed $V/t=-2$ and $n=0.75$ (see
  Fig.~\ref{fig:TransitionDensity}) with $L=128$. Fits (filled
  circles) using Eq.~(\ref{eq:fitentropy}) are quite accurate,
  allowing for the determination of the central charge $c$.}
\label{fig:entropy}
\end{figure}

Another prediction from CFT concerns the behavior of the central
charge $c$ of the model at the transition line. The central charge
somehow measures the effective Ising degrees of freedom in the
low-energy physics. It is expected to be one in both regions (one
gapless bosonic modes) around the transition but exactly equal to
$3/2$ at the transition due to the emergence of the Ising criticality
with central charge $c=1/2$. A simple way to extract the central
charge with DMRG is to use the behavior of the von Neumann
entanglement entropy $S_{\text{vN}}(x)$ of a block of size $x<L$. It
is defined as
\begin{equation}\label{eq:vNentropy}
S_{\text{vN}}(x) = -\mathrm{Tr}[\rho(x)\ln\rho(x)]\;,
\end{equation}
where $\rho(x)$ is the reduced density matrix of the block. As has
been emphasized recently in several studies, the use of the
entanglement entropy can provide crucial information for condensed
matter studies since it allows to detect quantum phase transitions
without any knowledge on the order parameters~\cite{Amico2008}.  It is
straightforwardly computed with the DMRG algorithm from the
eigenvalues of the reduced density matrix which is obtained at each
iteration. Following the ideas developed in
Refs.~\cite{Laflorencie2006,Sorensen2007}, a subleading oscillating
term emerges due to open boundary conditions. Similarly to what
happens in the XXZ model, we expect the oscillations to be related to
the local kinetic energy $t(x)$ (similar to the dimerization term).
Indeed, we can carry out fits in both regions by using the following
ansatz
\begin{equation}
\label{eq:fitentropy}
S_{\text{vN}}(x) = \frac{c}{6}\ln d(x|L) + A + B(t(x) - \overline{t})\;,
\end{equation}
where $\overline{t}$ is the mean value of $t(x)$ in the bulk, and
$A,B$ are two constants. The behavior of the entanglement entropy thus
gives access to the central charge $c$. On Fig.~\ref{fig:entropy}, a
clear jump of $c$ is observed at the transition. The expected value
$c=1$ is well reproduced in the two regions away from the critical
point. At the transition, the fit is not as accurate but restricting
it to the bulk region yields $c=1.51$, at the price of describing less
well the strong oscillations close to the edges (a fit including all
data yields $c=1.71$ but overestimates the behavior in the bulk). Note
that the singular behavior of $c$ differs from the continuous behavior
of $K$ and the gaps at the transition.  Note also that the $2k_F$
oscillations in the ADW/MS region are reminiscent of similar features
seen in local density and kinetic energy, and compatible with a $2k_F$
soft mode (see Fig.~\ref{fig:Evsk}a)~\cite{Legeza2007}. A last remark
is that the local kinetic energy oscillations on the transition line
should follow $t(x) \sim \cos(2k_Fx)x^{-(K+1)/4}$, i.e. an exponent
between those in the neighboring BCS/MDW region (provided $K>1/3$) and
ADW/MS region.

\section{Experimental signatures of MS phase}
\label{sec:experiments}

This section is devoted to the study of effects particularly relevant
for experimental set-ups. In addition to these results, we have
already discussed in Sec.~\ref{sec:gaps} the excitation gaps, relevant
for radio frequency spectroscopy, and the molecule fraction, which could be
measured experimentally from pictures resolving the hyperfine states.

\subsection{Effect of temperature on the $2k_F$ peak}

\begin{figure}[t]
\centering
\includegraphics[width=0.85\columnwidth,clip]{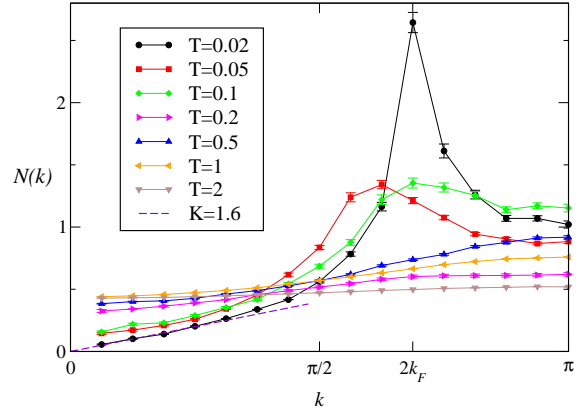}
\caption{SU(4) model: $N(k)$ obtained from QMC SSE simulations at
  various temperatures (in units of $t$) for $U/t=-2$, $L=32$ and
  $\mu=-3.4 t$. At low temperature, the small-$k$ linear behavior
  allows to extract $K\simeq 1.6$ which is compatible with DMRG
  estimate of 1.5 for same parameters. Moreover, the temperature
  effect allows to estimate at which energy scale the $2k_F$ peak
  appears. $k_F$ shown on the plot corresponds to the mean density
  $\langle{n}\rangle=1.4$.}
\label{fig:NktempSU4}
\end{figure}

Using QMC, it is possible to investigate the energy scale at which the
zero-temperature features become relevant. In
Fig.~\ref{fig:NktempSU4}, we study the effect of temperature on the
density structure factor $N(k)$ calculated on the SU(4) line for
different temperatures. Let us remind that we work in a
grand-canonical ensemble so that the density $\langle{n}\rangle$
varies with temperature (typically between 1.2 and 1.6 for this plot).
In particular, we have shown the $2k_F$ location corresponding to the
low-temperature density ($\langle{n}\rangle=1.4$). The two properties
of interest are the emergence of the $2k_F$ peak, and the linear
behavior at small-$k$. We observe that, below a typical temperature of
order $0.1t$, these two features qualitatively approach their $T=0$
behavior, while a quantitative estimate requires a much lower
temperature of order $0.02t$. If these features could be measured
experimentally, one could identify the two main regions from
the strength of the $2k_F$ and $4k_F$ peaks.

\subsection{Effect of the trap}

The effect of the trap confinement on two-component fermionic gases
loaded in optical lattices has been studied for
repulsive~\cite{Rigol2003} and
attractive~\cite{Xianlong2007,Molina2007} interactions. Trapped
fermions with attractive interactions were also studied for imbalanced
populations~\cite{Feiguin2007}, i.e. with no SU(2) symmetry.  The
Hamiltonian term corresponding to the harmonic confinement is:
\begin{equation}
\frac{\omega^2} 2 \sum_i (i - (L+1)/2)^2 n_i \;,
\end{equation}
with $\omega$ the trap frequency and $(L+1)/2$ the middle of the
chain. We thus take the box width $L$ to be larger than the cloud's
width not to induce boundary effect from the edges of the box. Without
a lattice, the chemical potential reads $\mu = \omega N_f/N$ for free
fermions. The thermodynamical limit is understood as taking the
$\omega \rightarrow 0$ limit while keeping $N_f \omega$ constant so
that the density at the center of the trap remains constant.
Consequently, $N_f \omega$ is similar to an effective density of the
system and, depending on it, several regimes are identified.

\begin{figure}[t]
\centering
\includegraphics[width=0.9\columnwidth,clip]{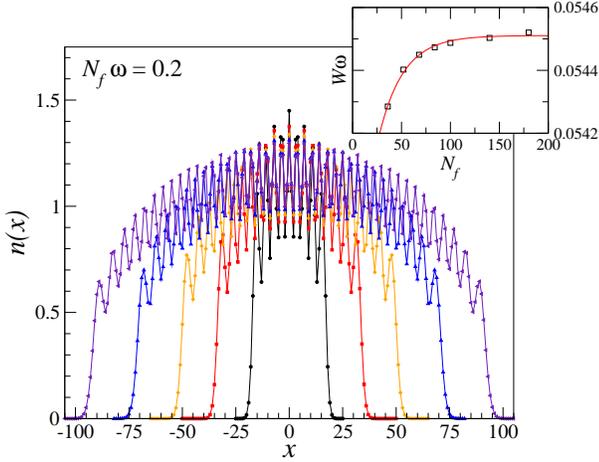}
\caption{Typical density profiles of the trapped condensate for
  fixed ``density'' $N_f \omega=0.2$ and $U/t = -4$, with respectively
  $N_f = 36,68,100,140,180$. The inset shows the scaling of the
  $W\omega$ in the thermodynamical limit, which well agrees with the TG
  behavior.}
\label{fig:fixedNomega}
\end{figure}

The density profile of the condensate $n(x)$ (or $n_i=n(x_i)$ in case
of a lattice) is directly accessible from experimental pictures. For
free fermions without an optical lattice in the Tonks-Girardeau (TG)
regime with $N$-color~\cite{Petrov2000,Dunjko2001,Tokatly2004}, one
has
\begin{equation}
\label{eq:TGprofile}
  n_{\text{TG}}(x) = n_0 \sqrt{1 - x^2/R_{\text{TG}}^2}
\end{equation}
with $R_{\text{TG}} \sim \sqrt{N_f/N\omega}$ and $n_0 \sim
\sqrt{N_f\omega/N}$, on which we observe that $n_0$ is kept constant
in the thermodynamical limit. These results are valid if the trap
evolves smoothly enough such that the local density approximation
(LDA) is expected to be a reasonable assumption. In presence of an
optical lattice, the energy per particle and the dispersion relations
are changed.  In this situation, LDA gives a density profile of the
type $ n(x) = n_0 \arccos(x^2/R^2-b^2)/\arccos(-b^2)$ for $|x| \leq
R\sqrt{1+b^2}$~\cite{Paredes2004}. The typical width of the density
distribution, which can be measured, is defined as $W = 2\sqrt{
  \frac{1}{N_f} \sum_i (i-i_0)^2 n_i}$. In the Tonks-Girardeau regime,
we have $W \sim R_{\text{TG}} \sim \sqrt{N_f/N\omega}$, so that $W
\sim \omega^{-1}$ in the thermodynamical limit. We found a similar
behavior for the spin-3/2 fermion model under study with a scaling
which agrees well with the TG one, as one can infer from the results
of Fig.~\ref{fig:fixedNomega}.

\subsubsection{Atomic density waves}

\begin{figure}[t]
\centering
\includegraphics[width=0.85\columnwidth,clip]{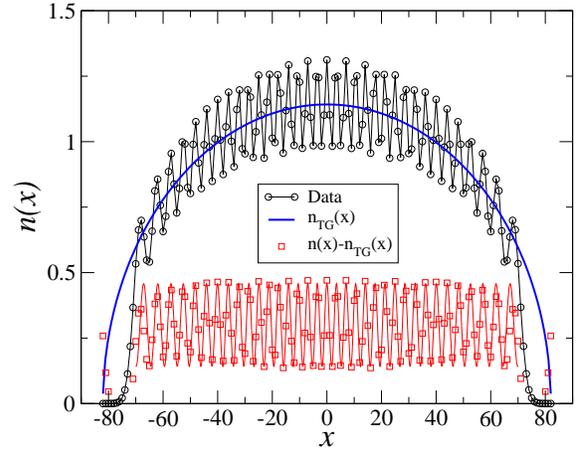}
\caption{Typical density profile of the trapped condensate on the
  SU(4) line ($U/t=-4$) with a smooth trap ($N_f \omega = 0.2$ and
  $N_f=140$). One observes that $2k_F \simeq \pi/2$ oscillations develop
  at the center of the cloud. Close to the edges, the wave-length of the
  oscillations increases, qualitatively following the local decrease
  of the local averaged density. A reasonable fit for the bulk physics
  is obtained using the Tonks-Girardeau result plus an oscillating
  term.}
\label{fig:illustration}
\end{figure}

For sufficiently smooth traps, the bulk of the condensate features the
typical $2k_F$ oscillations reminiscent of the ADW phase encountered
with open-boundary conditions. For instance,
Fig.~\ref{fig:illustration} shows a typical density profile on the
SU(4) line. The density profile can be reasonably fitted up to the
edges by a Tonks-Girardeau profile (Eq.~(\ref{eq:TGprofile})) plus an
oscillating term
\begin{equation}
 n(x) = n_{\text{TG}}(x) + \delta n \cos(2 {\tilde k}_F(x) x) \;,
\end{equation}
with the effective Fermi wave-vector
\begin{equation}
 {\tilde k}_F(x) = \frac{\pi}{4} n_0 \sqrt{1-x^2/({\tilde R}_{\text{TG}})^2} \;.
\end{equation}
In Fig.~\ref{fig:illustration}, we first fit the TG profile and
subtract it from the data to only keep the oscillating term that we
fit using the same value of $n_0$. The slight dependence of ${\tilde
  k}_F$ on $x$ accounts for the increase of the wave-length as the
density decreases towards the edges of the condensate. However, if
${\tilde R}_{\text{TG}}$ is a free parameter of the fit, we find that
${\tilde R}_{\text{TG}} \simeq 2 R_{\text{TG}}$ gives a better fit of
the oscillations. This discrepancy could be due to finite size
effects, as the TG profile should be valid for large enough systems
and far enough from the edges of the condensate. The condensate has
sharp edges (LDA usually fails to explain the behavior close to the
edges) and in the following, we call $a$ the radius at which the
density vanishes ($a \lesssim R_{\text{TG}}$).

\begin{figure}[t]
\centering
\includegraphics[width=0.9\columnwidth,clip]{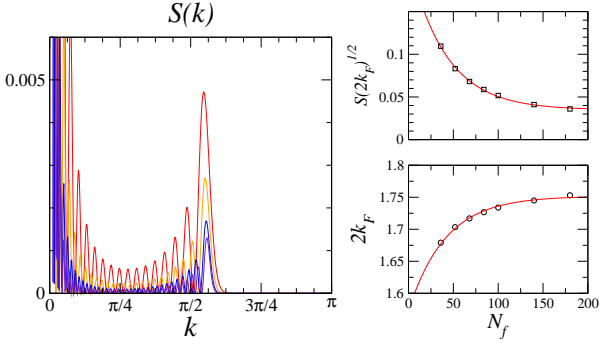}
\caption{The density structure factor $S(k)$ corresponding to
  Fig.~\ref{fig:fixedNomega} (same color code), and the scaling
  of the $2k_F$ peak
  position and amplitude in the thermodynamical limit.}
\label{fig:NkfixedNomega}
\end{figure}

The main question is now to discuss the thermodynamical limit of the
oscillations amplitude in the bulk $\delta n$. For non-trapped gases
in a box, we expect the oscillations to be zero in the middle of the
system (except for a translationally breaking phase) as the Friedel
oscillations decay away from the boundaries. In the SU(2) case, a
finite $\delta n$ has been found around the commensurate density
$n_0=1$~\cite{Xianlong2007,Molina2007}. These atomic-density waves
will have clear signatures in the density structure factor that can be
measured with light-scattering diffraction. The latter is defined by
\begin{equation}
S(k) = \Big\vert \frac{1}{N_f} \sum_j e^{ikj} n_j \Big\vert^2\;,
\end{equation}
and an example is given in Fig.~\ref{fig:NkfixedNomega}. The main
features of the structure factor is the peak at $2k_F \simeq \pi n_0
/2$ which signals the ADW oscillations of the trapped systems.
Oscillations in $S(k)$ are due to the finite size $2a$ of the
condensate and may vanish for large systems. We expect the height of
the peak to be proportional to $(\delta n)^2$ for large enough
systems.  The insets of Fig.~\ref{fig:NkfixedNomega} show the
evolution of $\sqrt{S(2k_F)} \sim \delta n$ and $2k_F$ (defined as the
$k$ at which the peak has its maximum) as a function of $N_f$.  Fits
are obtained for both quantities with an exponential law $s_0 + s_1
e^{-N_f/\xi}$. We infer from these results that $\delta n$ is finite
in the thermodynamical limit for our choice of parameters, similarly
to what was found in the SU(2) case~\cite{Xianlong2007}.

\subsubsection{Correlations and the BCS/ADW transition}

\begin{figure}[t]
\centering
\includegraphics[width=0.85\columnwidth,clip]{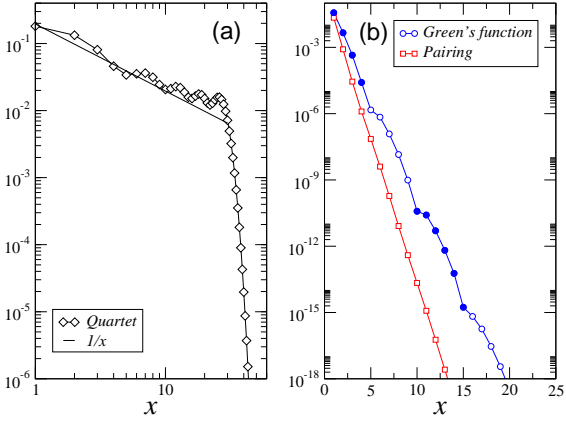}
\caption{Correlations in a trap system with $\omega = 0.002$, $U=-4$,
  $V=0$ and $N_f=36$. (a) Quartet correlations $Q(x)$ showing dominant
  MS correlations (from comparison with $1/x$) (b) Pairing and Green's
  function are short-range. The full (open) circles account for the
  +(-) sign coming from the $\sin(k_F x)$ term.}
\label{fig:TrappedCorrelations}
\end{figure}

\begin{figure}[t]
\centering
\includegraphics[width=0.95\columnwidth,clip]{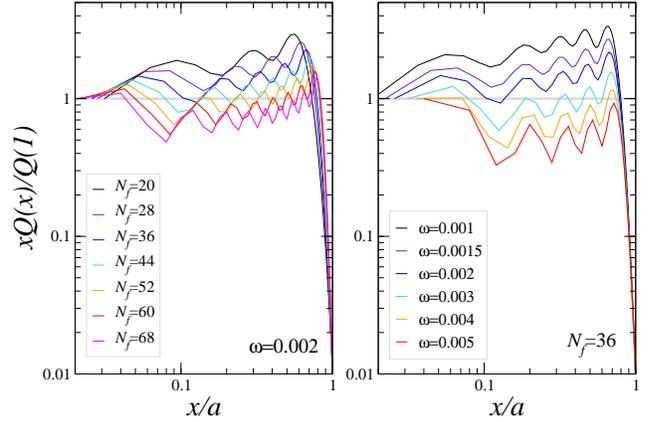}
\caption{(a) Correlations in a trapped system with $U=-4$ and $V=0$
  (a) for fixed $\omega = 0.002$ and various $N_f$. The distance $x$
  is rescaled by the trap radius $a$. (b) The same for fixed $N_f =
  36$ and varying $\omega$.}
\label{fig:TrappedCrossover}
\end{figure}

We now turn to the behavior of the correlations in trapped gases.
They are computed by fixing one point of the correlator at the center
of the cloud and letting the position $x$ of the other ranging from
the middle to the edge of the condensate. For a trapped condensate, a
gap computed from energy differences can be spoiled by edges effect.
To check the gapped nature of the one- and two-particle excitations
along the SU(4) line, we compute the Green's function and the pairing
correlation. From the behavior of the quartet correlations, we can
deduce the leading fluctuations in the bulk of the cloud. These
results are given in Fig.~\ref{fig:TrappedCorrelations} in which we
see that for this small value of the effective density ($N_f\omega =
0.07$), the physics in the bulk is essentially the same as for the
non-trapped system. Quartet correlations slightly increase as one
approaches the edge of the condensate. Assuming the LDA approximation
to work in such small systems, we expect the local Luttinger
parameter~\cite{Gangardt2003,Kollath2004} controlling the power-law
decrease of the correlations $K(x) = K(n(x))$ to increase with
decreasing densities, following the homogeneous system result of
Fig.~\ref{fig:KSU4}. As the quartet correlations behave as $2/K$ and
because $K$ decreases with density, they are actually reinforced by the
harmonic confinement. Trapping thus favors the observation of the MS
phase. By lowering the effective density $N_f\omega$, we expect a
crossover from sub-leading to leading quartet correlations. In
Fig.~\ref{fig:TrappedCrossover}, we show both the effect of changing
the number of particles $N_f$ while keeping $\omega$ constant and the
opposite situation where $\omega$ varies. We find that quartet
correlations always decreases slower than $1/x$ (corresponding to the
critical value $K_c=2$) below approximately $N_f\omega \simeq 0.08$.

\subsubsection{Effect of varying interactions and deep trap physics}

\begin{figure}[t]
\centering
\includegraphics[width=0.98\columnwidth,clip]{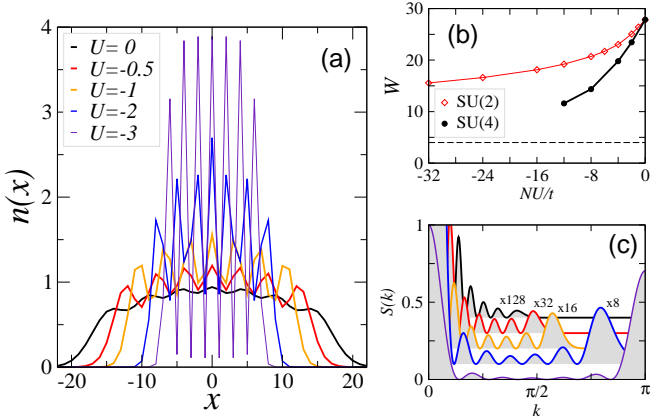}
\caption{(a) Typical density profiles of the trapped condensate for
  various interaction strength $U$ on the SU(4) line. There are $N_f =
  28$ fermions and $\omega = 0.05$. (b) Width of the condensate as a
  function of $U/t$. Its width is roughly divided by a factor 2.5. The
  SU(2) line has been computed with the same $\omega$ but with 14
  fermions. The width is much more sensitive to $U$ in the SU(4) case.
  (c) Structure factors have been rescaled for clarity.}
\label{SU4Trapped:collapse}
\end{figure}

In this section, we address the question of the effect of varying
interactions on the density profile of the condensate. We expect the
width of the condensate to strongly depend on interactions: repulsive
interactions make the condensate inflate while attractive interactions
can strongly reduce $W$. In the large trap frequency limit, we
furthermore have a minimal width of the condensate due to Pauli's
exclusion principle which depends on the number of colors:
$W^{\text{min}} = \sqrt{ \frac{1}{3}\big((N_f/N)^2 - 1 \big)} \sim
N_f/\sqrt{3}N$. Thus, starting from free electrons and keeping
$N_f\omega/N$ constant, the ratio $W^{\text{min}}/W^{\text{free}} \sim
\sqrt{N_f\omega/N}$ should be constant. We give a comparison with the
evolution of the width in the SU(2) case as a function of $NU$: the
collapse of the condensate is faster in the SU(4) case. In
Fig.~\ref{SU4Trapped:collapse}, we show, for a constant and rather
large effective density $N_f\omega = 1.4$, the evolution of the
condensate when increasing (negative) $U$ along the SU(4) line. Two
effects are observed: the width of the condensate strongly decreases
which induces a sharp increase of the density at the center of the
trap.  Consequently, the effective Fermi wave-vector increases, which
can be observed in the density structure factor. Similar effects have
been found for the SU(2) attractive
model~\cite{Machida2006,Xianlong2007a}.  Because the effective density
is large enough, the commensurate ADW$^{\pi}$ phase develops when $n_0
\geq 2$. A strong signal at $k=\pi$ is then present in the density
structure factor.

\begin{figure}[t]
\centering
\includegraphics[width=0.95\columnwidth,clip]{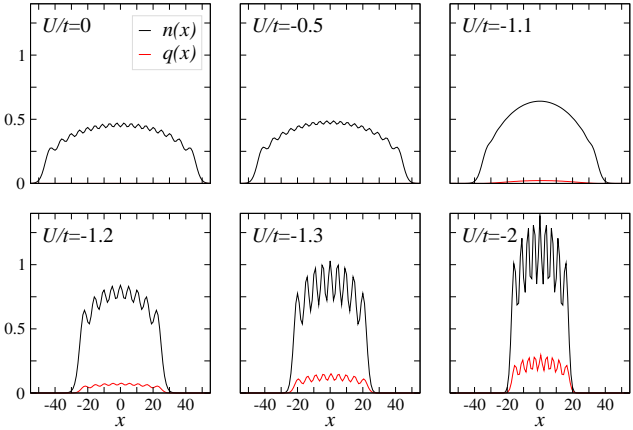}
\caption{Evolution of the profiles along the $V=-2$ line with decreasing
 $U$ showing the crossing of the transition (located at $U/t=-1.2$).}
\label{fig:TransitionProfiles}
\end{figure}

Moving away from the SU(4) line, we can investigate the behavior of
the density profile across the transition between the two regions of
the phase diagram of Fig.~\ref{fig:phase-diagram}. Indeed, we have
seen in Sec.~\ref{sec:TransitionDensity} that the transition line
hardly depends on the density. Therefore, for inhomogeneous density
profiles, we expect the transition to hold for identical parameters,
and that the whole cloud will be affected by the phase transition
(meaning that no domains will form). Note that this is not true for
the crossover lines inside each region of
Fig.~\ref{fig:phase-diagram}, and that is seen through the variation
of $K(n(x))$ in Fig.~\ref{fig:TrappedCrossover}.  The typical
evolution of $n(x)$ along the $V/t=-2$ line (varying $U$) is given in
Fig.~\ref{fig:TransitionProfiles}. Two main effects are visible: the
width of the condensate is roughly divided by two, and the amplitude
of the Friedel like oscillations are strongly enhanced in the bulk.
Remarkably, because of the strong increase of the density at the
center of the cloud in the ADW/MS region, the wave-vector of the
oscillations hardly changes (it is $4{\tilde k}_F$ in the BCS phase,
but $2{\tilde k}_F$ in the ADW/MS region). If one is able to measure
the local density of molecules, a crossover from zero to a significant
value will be also observed across the transition, similarly to what
was found in Fig.~\ref{fig:TransitionDensity}. The fact that the
density of quartets is spread over the whole condensate through the
transition line supports the absence of domain formation.

\begin{figure}[t]
\centering
\includegraphics[width=0.85\columnwidth,clip]{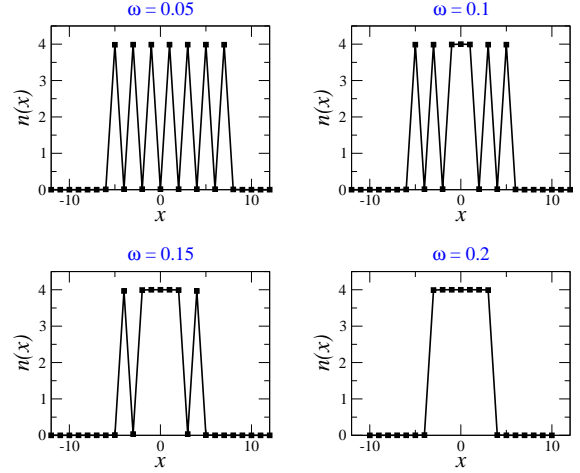}
\caption{Starting with a system with strongly bound quartets
  ($N_f=28$, $U/t = -8$), one can increase the frequency of the trap.
  Quartets have a tendency to repel each other but when the trap is
  too deep, they progressively melt at the center of the trap.}
\label{fig:meltingSU4}
\end{figure}

If the trap is very deep, corresponding to large values of the
effective density $N_f\omega$, all atoms will form an homogeneous
condensate in the middle of the trap, the width of which is
$W^{\text{min}}$ as discussed above. As displayed in
Fig.~\ref{fig:meltingSU4}, this state emerges from the melting of the
ADW$^{\pi}$ phase which can be qualitatively understood as the mere
competition between the effective nearest neighbor repulsion of the
quartets and the potential energy of the trap. This effect is very
similar to the one found for the SU(2) model in
Ref.~\cite{Xianlong2007a}. For $\omega = 0.05$, the density profile is
shifted from the center of the trap. Actually, the energy of such a
shifted state is much smaller than other energy scales (equal to
$0.00875t$ within a classical approximation), so that it is nearly
degenerate to the ground-state and DMRG gets locked into it because
the effective hopping term of the molecules is too small (for large
$\vert{U}\vert$).

\subsection{Flux quantization}

In standard electronic systems, flux quantization experiments can
directly measure the electric charge of the carriers by considering a
\emph{ring} geometry threaded by a magnetic flux and look at the flux
periodicity of the total energy. In the case of neutral cold atoms,
such a flux analogy could be realized thanks to the possibility of
rotating the trap. Therefore, it could be possible to prove the
existence of $N$ particle bound states by checking if minima of the
energy are degenerate~\cite{Lecheminant2005}.  Moreover, ring-shape
geometries have been realized experimentally
\cite{Sauer2001,Heathcote2008,Ryu2007} so that such experiments could
be performed in the near future.

\begin{figure}[t]
\centering
\includegraphics[width=0.85\columnwidth,clip]{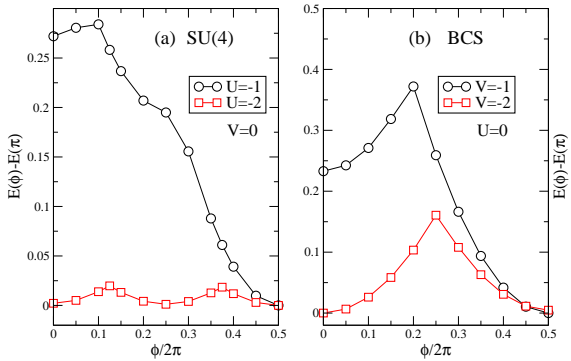}
\caption{Energy vs flux for a periodic chain of length $L=8$ with 8
  particles for various interactions. Energies are measured relative
  to the absolute ground-state energy (here at $\phi=\pi$). (a) SU(4)
  model showing the appearance of 4 minima; (b) BCS phase with 2
  minima. Note that only flux between 0 and $\pi$ are shown since data
  are symmetric $E(-\phi)=E(\phi)$ and $2\pi$ periodic.}
\label{flux.fig}
\end{figure}

We have performed exact diagonalization for the SU(4) case on small
chains of length $L=8$. Although these sizes are relatively small, we
expect that bound-state formation can already be checked since it is
a local process. Indeed, as the interaction strength increases, we see
that the number of minima changes from one to four (see
Fig.~\ref{flux.fig}(a)) in the ADW/MS phase, while it increases from
one to two in the BCS/MDW phase (see Fig.~\ref{flux.fig}(b)), in full
agreement with the predictions of the low-energy
approach~\cite{Lecheminant2005}. These observations are compatible
with our predictions of four- and two-particle bound states,
respectively.  Note that the overall energy scales decrease since the
band width is proportional to the effective molecular or pair hopping.
Similar calculations suggesting pairing and superfluidity in a
multicomponent fermionic model have been proposed
recently~\cite{Wang2007}. As a final comment, let us remind the reader
that the effect of such a flux corresponds to a twist in the boundary
conditions, and therefore vanishes in the thermodynamic limit where
$E(\phi)$ becomes flat.  Still, such an effect can be observed on
finite lattices.

\section{Conclusion}
\label{sec:conclusion}

Motivated by fermionic cold atoms experiments where generically many
hyperfine states coexist, we have investigated spin 3/2 fermions with
contact interactions in an optical lattice. We focus on the attractive
case for a generic density.  By using large-scale numerical
techniques, we describe the phase diagram and discuss all competing
phases. In particular, at low density, we confirm the existence of a
large molecular superfluid phase where dominant correlations are
superfluid-like made of four-particle bound-states. This phase has a
large extension and is not restricted to the SU(4) model. In another
region of the phase diagram, two-particle pairs become gapless,
leading to a BCS phase. We have shown that the phase transition
between MS and BCS phases can be located using entanglement
measurements, such as the von Neumann entropy or the molecule
fraction.

In order to make contact with possible experimental observations of
such phases, we have investigated the role of the trapping potential.
In many respects, correlations inside the bulk of the condensate are
similar to the homogeneous case if the effective density is low
enough. ADW oscillations can be probed from the density structure
factor. Furthermore, we give an estimate for the crossover effective
density below which leading MS fluctuations are dominant. Moreover,
playing with the trap can provide useful informations about the size
of the condensate and the density correlations, which are accessible
experimentally. For instance, deep in the ADW/MS phase, the physics
can be understood from tightly bound ``molecules'' that act as
hardcore bosons. These objects could be measured either by looking at
the molecules fraction or by using rf spectroscopy. Finally, we
propose to distinguish between MS and BCS phase by using the molecule
fraction or ring-shape geometries. We hope that such experiments will
be performed in the near future.


\begin{acknowledgement}
  We would like to thank T. Barthel, E. Boulat, A. M. L\"auchli, F.
  Heidrich-Meisner, P. Schuck, and S.~R. White for useful discussions.
  GR and SC thank IDRIS (Orsay, France) and CALMIP (Toulouse, France)
  for use of supercomputers.
\end{acknowledgement}

\end{document}